\documentclass[reprint,showpacs,preprintnumbers,prl,10pt,aps,groupedaddress]{revtex4-1}
\usepackage{graphicx}
\usepackage{dcolumn}
\usepackage{bm}
\usepackage{epsfig}
\usepackage{latexsym}
\usepackage{amsmath}
\usepackage{color}
\usepackage{array}

\newcommand{\<}{\langle}
\newcommand{\e}{\varepsilon}
\renewcommand{\>}{\rangle}
\renewcommand{\(}{\left(}
\renewcommand{\)}{\right)}

\begin{document}

\title{Superfluidity and dimerization in a multilayered system of fermionic
polar molecules} 
\date{\today}

\author{Andrew C. Potter$^1$}\author{Erez Berg$^2$}\author{Daw-Wei Wang$^3$}\author{Bertrand I. Halperin$^2$}\author{Eugene Demler$^2$}
\affiliation{$^1$Department of Physics, Massachusetts Institute of Technology,
Cambridge, Massachusetts 02139}
\affiliation{$^2$Department of Physics, Harvard
University, Cambridge, Massachusetts 02138}
\affiliation{$^3$ Physics Department and NCTS, National Tsing-Hua University, Hsinchu 30013, Taiwan}

\begin{abstract}
We consider a layered system of fermionic molecules with permanent dipole moments aligned perpendicular to the layers by an external field.  The dipole interactions between fermions in adjacent layers are attractive and induce inter-layer pairing.  Due to competition for pairing among adjacent layers, the mean-field ground state of the layered system is a dimerized superfluid, with pairing only between every-other layer.  We construct an effective Ising-XY lattice model that describes the interplay between dimerization and superfluid phase fluctuations.  In addition to the dimerized superfluid ground state, and high-temperature normal state, at intermediate temperature, we find an unusual dimerized ``pseudogap" state with only short-range phase coherence.  We propose light scattering experiments to detect dimerization.
\end{abstract}

\maketitle

The long-range and anisotropic nature of dipole-dipole interactions offers
new opportunities for ultra-cold polar molecules, beyond what is possible
for cold-atom systems with only short-range, isotropic contact interactions \cite{Reviews}.
A variety of exotic many-body states including $p_x+ip_y$ fermionic superfluids \cite{p-wave}, and nematic non-Fermi liquids \cite{Biaxial_Nematic}, are predicted to occur in cold dipolar systems.  Additionally, polar molecules could provide a robust toolbox for engineering novel lattice-spin Hamiltonians \cite{spin_toolbox} or hybrid devices for quantum information processing \cite{Hybrid_devices}. Recent progress towards trapping and cooling atoms and molecules with permanent electric or magnetic
dipole moments have opened the door to exploring these exotic states of
matter experimentally \cite{Experiments}. In order
to prevent the system from collapsing due to the attractive head-to-tail
part of the dipolar interaction \cite{BEC_collapse}, it has been proposed (\cite{Bilayer},\cite{Chains}) to create stacks
of dipolar particles confined to a set of parallel planes.

\begin{figure}[t]
\begin{center}
\includegraphics[width=2.9in]{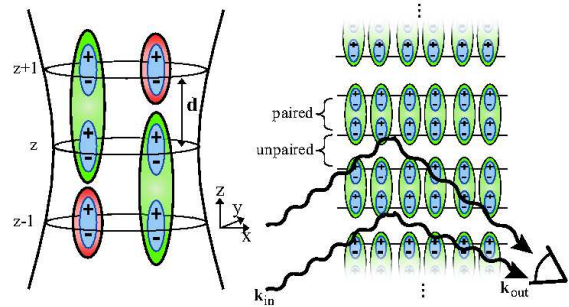}
\end{center}
\vspace{-.3in}
\caption{Schematic representation of competition for pairing among adjacent pairs of layers, including depiction of the optical confinement beam which creates the stack of 2D sheets (left), and illustration of one of the two equivalent dimerized pairing ground states for a many layered system (right).  The wavy lines illustrate the proposed light-scattering detection scheme discussed below in the text.}
\label{fig:System}
\end{figure}

In this Letter, we consider a stack of two-dimensional layers of
polar fermions whose dipole moments, $\vec{D}$, are
aligned along the stacking direction (z-axis) by an external field
(see Fig. \ref{fig:System}). The
dipole-interaction, $V_{d}=\frac{D^{2}}{r^{3}}\left( 1-3\frac{z^{2}}{r^{2}}%
\right) $, is purely repulsive between fermions in the same layer, and
partially attractive (for $r<  \sqrt{3}z$) between fermions in different
layers. The attractive interlayer component of the dipole interaction
induces BCS pairing between layers with adjacent layers competing for
pairing. We demonstrate that competition between adjacent layers favors
dimerization, with pairing only between even or odd pairs of layers (Fig. \ref{fig:System}).

We find three distinct phases: a high temperature disordered
phase, a fully ordered phase characterized by dimerized pairing amplitude
and quasi-long range ordered (QLRO) pairing phase in each layer (Fig.
1), and a dimerized ``pseudogap" phase with only short-range superfluid
correlations. The latter phase is particularly interesting, since
it can only be characterized by a composite \emph{four fermion}
dimerization order parameter. Therefore, this phase does not admit
a mean field (Hatree-Fock) description.
This is analogous to spin nematics \cite{spin_nematics}
and charge $4e$ superconductors \cite{charge4e}, which are both phases of
strongly interacting fermions that can only be characterized by
composite order parameters.

\par
\textit{Fermionic Pairing in a Layered System - } The action for
an $N$-layer system in terms of fermionic fields $\psi$ is
\begin{eqnarray} S &=& \sum_{z=1}^N\sum_{\mathbf{k}} \psi_{z,\mathbf{k}}^\dagger\(\partial_\tau+\e_\mathbf{k}-\mu\)\psi_{z,\mathbf{k}}\nonumber \\
&&\hspace{-.2in}-\sum_{z,z'=1}^N\sum_{\mathbf{k},\mathbf{k'},\mathbf{q}}
\psi_{z,\mathbf{k}'}^\dagger\psi_{z',\mathbf{q}-\mathbf{k}'}^\dagger
V_{|\mathbf{k}-\mathbf{k}'|}^{(z,z')}
\psi_{z',\mathbf{q}-\mathbf{k}}\psi_{z,\mathbf{k}} \label{eq:SFermionic}\end{eqnarray}
where $z$ and $z'$ are (integer) layer labels,
$\psi^\dagger_{z,\mathbf{k}}(\tau)$ creates a fermion with
in--plane momentum $\mathbf{k}$ and imaginary time $\tau$
in layer $z$. (The $\tau$ labels have been suppressed above.)
$V^{(z,z')}_{q}$ is the dipolar interaction
between layers $z$ and $z'$,\newpage \noindent Fourier transformed with respect to
the in-plane separation, for example: $V^{(z,z\pm 1)}_q = -D^2 qe^{-qd}$.

By solving the BCS gap equation: $\Delta_{z,\mathbf{k}} =-\sum_{\mathbf{k}'}
V^{(z,z+1)}_{\mathbf{k}-\mathbf{k}'}\<\psi_{z+1,-\mathbf{k}'}\psi_{z,\mathbf{k}'}\>$, we find that the attractive interlayer interactions induce fermionic pairing between adjacent layers $z$ and $z\pm 1$ (see supplement for details). Interaction between
next-nearest layers and beyond is small, and will be neglected
throughout most of this Letter. To decouple the four-fermion
interaction term, we introduce Hubbard-Stratonovich (H-S) fields
$\Delta_z(\mathbf{r}_\perp)$ associated with the pairing order
parameters (where $\mathbf{r}_\perp$ is the in--plane coordinate),
and integrate out the fermionic degrees of freedom\cite{comment-phase}. Expanding the
resulting fermionic determinant to quartic order (valid in the
vicinity of the phase transition where $|\Delta|$ is small) yields
the following Ginzburg-Landau (GL) free energy:
\begin{eqnarray} F &=& \sum_z \int d^2r \Big{(} \kappa|\nabla_{\hspace{-.05in}\perp}\Delta_z|^2+r|\Delta_z|^2 \nonumber \\ &&\hspace{.5in}+u|\Delta_z|^4+2u|\Delta_z|^2|\Delta_{z+1}|^2\Big{)} \label{eq:GL}\end{eqnarray}
where $\nabla_{\hspace{-.05in}\perp}$ denotes the gradient
restricted to the xy-plane. The GL coefficients are given by $\kappa=\frac{7\zeta(3)}{32\pi^3}\frac{\e_F}{T^2}$, $r=\nu t$, and $u=\frac{1.7}{32}\frac{\nu}{T^3}$
where $\zeta$ is the Riemann zeta-function,
$t=(T-T_c)/T$ is the reduced temperature, $\e_F$ is the
Fermi-energy, and $\nu$ is the two-dimensional density of states (for details we refer the reader to the supplement).

An important feature of this free energy is that the H-S expansion
does not generate $|\partial_z\Delta|^2$ terms, but only terms of
the form $\partial_z|\Delta|^2$.  The absence of
$|\partial_z\Delta|^2$ terms is not an artifact of the H-S
expansion; rather, it is guaranteed by particle number
conservation for each layer individually.  Particle conservation
for each layer stems from the absence of interlayer tunneling, and
formally corresponds to $N_{\text{Layers}}$ independent $U(1)$
phase rotation symmetries, $\psi_{z}\rightarrow
e^{i\theta_{z}/2}\psi_{z}$, of fermion fields $\psi_{z}$ in layer
$z$.  In contrast to other quasi-two-dimensional systems, such as
superconducting thin films where the behavior of the system tends
towards three-dimensional as the film thickness is increased,
two-dimensional Berezinskii-Kosterlitz-Thouless (BKT) physics remains important even for a large
number of layers.
\par
\textit{Mean-Field Ground State - } The
$2u|\Delta_i|^2|\Delta_{i+1}|^2$ term in (2) indicates that
adjacent pairs of layers compete with each other for pairing. For
$N_{\text{layers}}>3$, the mean-field theory predicts that it is
energetically favorable for the system to spontaneously dimerize
into one of two equivalent configuration where $\Delta$ vanishes
between every-other layer: $|\Delta_j| =
\frac{1}{2}\left[1\pm(-1)^j\right]\Delta_0$ (see Fig.1). The situation for
$N_{\text{layers}}=3$ is more subtle, and we defer its discussion.

\textit{Effective Lattice Model for Many-Layer System - }  The
above mean-field analysis suggests that the relevant degrees of
freedom for a many-layer dipolar system are Ising-like
dimerization between even or odd layers, and two-dimensional
XY-like phase fluctuations of the interlayer pairing
order-parameters.  In order to describe phase transitions in this
system, we course-grain the GL theory (in-plane) over length-scales below the
GL coherence length $\xi_{\mathrm{GL}}\equiv (\kappa/|r|)^{1/2}$,
and obtain the following effective lattice model \cite{XY_Ising}
\begin{eqnarray} F &=& \sum_{z}\Big{\{}K_z\sum_i \sigma_{z,i}\sigma_{z+1,i} -K_\perp\sum_{\<ij\>}\sigma_{z,i}\sigma_{z,j} \nonumber\\
&&\hspace{.1in}-\sum_{\<ij\>}J(\sigma_{z,i},\sigma_{z,j})\left[\cos\(\theta_{z,i}-\theta_{z,j}\)-1\right]
\Big{\}} \label{eq:LatticeModel}\end{eqnarray}  of Ising
variables $\sigma_{i,z}\in\{\pm 1\}$ coupled to XY phase-variables
$\theta_{z,i} = \text{Arg}\Delta_{z}(\vec r_i) \in [0,2\pi]$ where $z$ labels physical layers, $i$
labels lattice sites in the xy-plane, and
$J(\sigma_{z,i},\sigma_{z,j})\equiv
{J_0}\(1+\sigma_{z,i}\)\(1+\sigma_{z,j}\)/4$. \par
\begin{figure}[ttt]
\begin{center} \hspace{-.2in}
\includegraphics[width=3.2in]{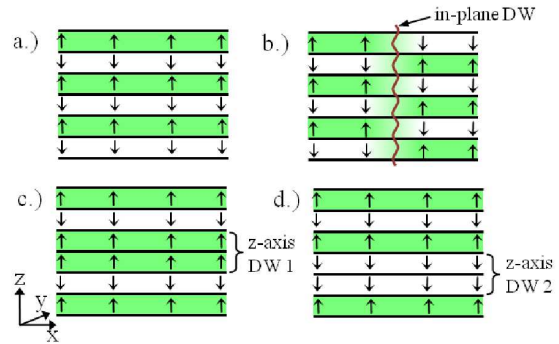}
\end{center}
\vspace{-.3in}
\caption{Schematic depiction of fully dimerized phase (a), in-plane Ising domain-wall (DW) (b), z-axis Ising DW with pairing between two adjacent pairs of layers (c), and z-axis Ising DW with no-pairing for two adjacent pairs of layers (d).  Green shading between layers indicates pairing (color online).}
\label{fig:DW}
\vspace{-.2in}
\end{figure}

    In the lattice model, $\sigma_{z}=+1$ ($\sigma_{z}=-1$) indicates that layers
$z$ and $z+1$ are paired (un-paired respectively). The uniformly
dimerized ground state of the multilayer system corresponds to
anti-ferromagnetic Ising order along the z-axis and ferromagnetic
order within the xy-plane. Ising domain walls correspond to
regions where pairing switches between the two equivalent
dimerization configurations over a distance of the order of the GL
coherence length, either along the z-axis or within the xy-plane.
The coupling constants $K_z$ and $K_\perp$ reflect the energy cost
of deforming the magnitude of the pairing order parameter,
$|\Delta|$, to form a domain wall along the z-axis or in the
xy-plane respectively (see Fig. \ref{fig:DW}). \par

The coupling $J(\sigma_{z,i},\sigma_{z,j})$ corresponds to the
average superfluid stiffness $\rho \sim \kappa |\Delta|^2$ in the
vicinity of lattice site $(i,z)$ and determines the energy cost of
twisting the phase of the order parameter, $\theta_{z,i}$, between
sites $i$ and $j$ in the same plane. The local stiffness is
non-zero wherever $\sigma_{z,i}=+1$, and zero otherwise\cite{comment-phase}.
\par
The lattice model couplings ($K_{z},K_\perp, J_0$) can be estimated from the GL model.  An in-plane dimerization
domain wall along the x-direction (Fig. \ref{fig:DW}b)
corresponds to pairing configurations of the form $\Delta_{z}(x) =
\frac{\Delta_0}{2}[1+(-1)^z\alpha(x)]$ where $\Delta_0^2 =
\frac{|r|}{2u}$
, and $\alpha(x)$ is a function that changes from $-1$ to $+1$
around $x=0$, and tends to a constant away from $x=0$. 
Minimization of the free energy with respect to $\alpha(x)$ yields
$\alpha(x)=\tanh\(2x/\ell_{\mathrm{DW}}\)$ where
$\ell_{\mathrm{DW}}\equiv
\sqrt{\frac{32\kappa}{3r}}$\cite{comment-phase}. The corresponding
free energy cost per unit length is $U_{DW}^{(\perp)} =
\ell_{\mathrm{DW}} \frac{|r|^2}{8u}$.

There are two possible z--axis domain wall configurations, shown
in Fig. \ref{fig:DW}c,d. To determine their free energy cost, we
consider a system with periodic boundary conditions along the z
axis, and compare the free energy of the ground state to that of
the domain wall configurations.  This yields an energy cost per
unit area $U_{DW}^{(z)} = \frac{|r|^2}{8u}$ for both types of
domain walls.  Setting the lattice spacing equal to $\ell_{DW}$,
the energetics of in-plane and z-axis domain walls are reproduced
by  $K_z = 2K_\perp =\frac{4\kappa|r|}{3u}$. In order to determine the lattice phase stiffness $J_0$, we equate the cost of an infinitesimal phase twist, $\theta_{z,j}=\theta_{z,i}+\delta\theta$, in a fully paired layer ($\sigma_z=1$) to the corresponding cost in the GL free energy (Eq. \ref{eq:GL}). This gives $J_0=\frac{\kappa|r|}{u}$.

\begin{figure}[ttt] \hspace{-.15in}
\includegraphics[width = 3.3in]{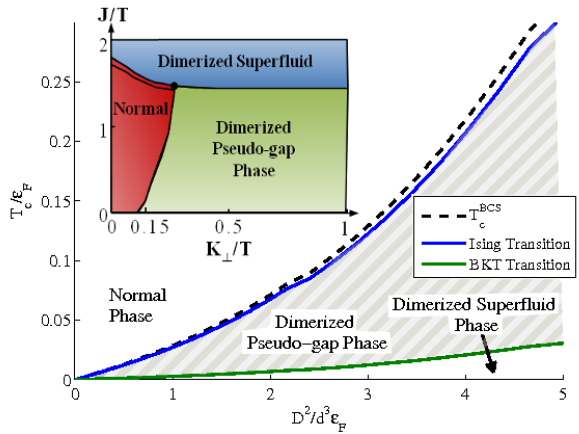}
\vspace{-.1in}
\caption{The lattice model phase-diagram, calculated using the temperature dependence of the GL coefficients in (3) and plotted in terms of dipole interaction strength $D^2/d^3$ and temperature T, each measured in units of $\e_F$ (main figure).  Phase diagram predicted by the effective lattice model for generic model parameters, with $K_z/K_\perp=2$. Double line indicates first-order transition (inset).}
\label{fig:PhaseDiagram}
\vspace{-.2in}
\end{figure}
    The lattice model (Eq. \ref{eq:LatticeModel}) describes three-dimensional Ising spins coupled to many independent two-dimensional XY layers.  For temperatures near or below the Ising transition temperature, the Ising variables have large correlation lengths and hence see an average over many independent layers of XY spins.  With this self-averaging property in mind, we decouple the XY and Ising variables in a mean-field factorization
\begin{eqnarray}  F_{\sigma} = K_z\sum_{\<zz'\>,i} \sigma_{zi}\sigma_{z'i} -K_\perp^{(\text{eff})}\sum_{z,\<ij\>}\sigma_{zi}\sigma_{zj}-h\sum_{z,i}\sigma_{zi} \nonumber \end{eqnarray}
\vspace{-.25in}
\begin{eqnarray} F_{\text{XY}}= - \sum_{\<ij\>}\frac{J_0}{4}\left[1+(-1)^z\sigma_0\right]^2\cos(\theta_{z,i}-\theta_{z,j}) \nonumber \end{eqnarray}
where $K_{\perp}^{(\text{eff})} = K_\perp+\frac{J_0}{4}\(\frac{A+B}{2}\)$ and $A,B \equiv
\<\cos(\theta_i^{(e/o)}-\theta_j^{(e/o)})\>_{F_{\text{XY}}}-1$ are
the averages (with respect to $F_{\text{XY}}$) of the cosine terms
in even and odd layers respectively, $\sigma_0 \equiv
\<\sigma\>_{F_{\sigma}}$, and $h =
\(\frac{A-B}{2}\)\frac{J_0}{2}$.\par The decoupled Ising model and
XY models can then be analyzed separately but self-consistently.
A mean-field analysis is adequate for 3D Ising model. 
The phase action is treated by a variational self-consistent
harmonic approximation (SCHA) \cite{SCHA_for_XY}.  While the SCHA provides a
reasonable estimate of the location of the 2D BKT transition, it
spuriously predicts a strong first-order transition in which
$\<F_{\text{XY}}\>_{\text{SCHA}}$ drops abruptly to zero at the XY
transition temperature, $T_{XY}$. At higher temperatures, the SCHA
dramatically underestimates the contribution to energy density
from phase fluctuations.   In order to avoid this undesirable
feature, we supplement the SCHA value for
$\<\cos\Delta_{ij}\theta\>_{\text{SCHA}}$ with a
high-temperature expansion for $T>T_{XY}$:
\begin{equation} \<\cos\Delta_{ij}\theta^{(z)}\> = \left\{\begin{array}{ll} \<\cos\Delta_{ij}\theta^{(z)}\>_{\text{SCHA}} &; T<T_{XY} \\ J(\sigma_0,\sigma_0)/2T &; T>T_{XY}\end{array} \right.\end{equation}

Fig. \ref{fig:PhaseDiagram} shows the phase diagram predicted by the effective
lattice model.    The main figure displays the phase
diagram where the model parameters are taken from the GL
coefficients in (3).  The BCS transition temperature,
$T_c^{\text{BCS}}$, is obtained by solving numerically the BCS gap
equation for the dipole potential.  Whereas the dimerization transition
occurs close to the mean-field BCS transition temperature,
$T_c^{\text{BCS}}$, the BKT transition to phase QLRO occurs at a lower temperature, leaving an intermediate region with full dimerization but only short range superfluid correlations.

Recent experiments on 3D clouds of ultra-cold $^{40}\text{K}^{87}\text{Rb}$ molecules
have achieved densities on the order of $n_{3d} =
10^{12}\text{cm}^{-3}$ and permanent electrical dipole moments of
up to $0.566$ Debye \cite{Experiments}.  If similar densities
were achieved in a layered system with layer spacing on the order
of $400$nm, the ratio of typical dipole interactions to
Fermi-energy would be $D^2/(4\pi\e_0 d^3\e_F)\sim 3$.


While the GL parameters in Eq. \ref{eq:GL} provide an initial estimate of the lattice-model coupling constants,  in principle, the model coefficients can be renormalized by higher order terms in the GL expansion.  The inset shows the phase diagram for generic
values of the model parameters $K_\perp$ and $J_0$ with
$K_z/K_\perp=2$ (the qualitative features do not depend
sensitively this ratio).  An additional feature emerges for generic coefficients: for $J$ sufficiently bigger than $K$, there is a tri-critical point where the BKT and Ising transitions fuse into a weakly first-order phase transition.

\textit{Order parameter and detection -} The dimerized phase
breaks translational symmetry in the $z$ direction. It can
be characterized by the following \emph{four fermion} order
parameter: $\mathcal{D}=\<n_{z-1,r}n_{z,r}-n_{z,r}n_{z+1,r}\>$,
where $n_{z,r}=\psi^\dagger_{z,r}\psi_{z,r}$ is the local fermion
density. For finite transverse confinement, in the dimerized phase, every two paired
layers shift slightly towards each other. The displacement scales
as $\delta z\propto \Omega_z^{-2}$, where $\Omega_z$ is the layer--confinement frequency in the z--direction. The dimerized phase can
be detected by the appearance of new Bragg peaks in elastic light
scattering (see Fig. \ref{fig:System}) with wavevector
$\mathbf{Q}=n\pi\mathbf{\hat{z}}/d$, $n=1,3,\dots$, with
intensity $\sim \delta z^2$.

In the strong--confinement limit, $\Omega_z\rightarrow \infty$,
the particle density does not show any sign of dimerization.  However in this regime, the
dimerized phase could still be detected by measuring correlations between
the amplitudes of light scattered at different wavevectors: 
$\<n_{\mathbf{q}}n_{\mathbf{q'}} \>\propto
n_0^2\delta_{\mathbf{q}+\mathbf{q'}}+\delta_{\mathbf{Q}-\mathbf{q}-\mathbf{q'}}\mathcal{D}$,
where $\mathbf{q}$ and $\mathbf{q'}$ are two scattering
wavevectors and $n_0$ is a constant.
\par
\textit{Three Layer Case -} The three layer system is a special case that requires more careful analysis.  If one proceeds as above and includes interactions only between neighboring layers, the system possesses an extra $SU(2)$ symmetry generated by: $I^{z} = \int d^2r \( \psi^\dagger_3\psi_3 - \psi^\dagger_1\psi_1\)$ and $I^{\pm} = \int d^2r \( \psi^\dagger_3\psi_1\pm i \psi^\dagger_1\psi_3\)$.  The $U(1)$ generator $N_2 = \int d^2r \psi^\dagger_2\psi_2$ completes the $SU(2)$ symmetry to $U(2)$. These generators commute with ${\cal H} = {\cal H}_{\rm kin} + V_{12} + V_{23}$, where $V_{ij}$ is the interaction between layers $i$ and $j$.  This $U(2)$ symmetry dictates that, to all orders in the GL expansion, the free energy should be a function of $\(|\Delta_{1}|^2 + |\Delta_{2}|^2\)$ only, which does not energetically distinguish dimerization from uniform pairing.
\par
    However intralayer and next-nearest neighbor interactions $\tilde{V} = V_{13}+\sum_{j=1}^3 V_{jj}$ break the SU(2) symmetry of the three layer system, and generate a quartic term of the form $-|v| |\Delta_1|^2 |\Delta_2|^2$ in the GL free energy.  This term is relevant \cite{NLSM_RG} (in the renormalization group sense) and hence, we expect the trilayer system to exhibit uniform pairing with $|\Delta_1| = |\Delta_2|$.  In contrast, for $N_{\text{Layers}}>3$, already the dominant nearest neighbor interactions strongly favor dimerization and $\tilde{V}$ only produce small subleading corrections.

\textit{Discussion - }
    Our analysis of the layered dipolar Fermi system predicts a sequence of two phase-transitions: an Ising-like dimerization transition followed by a BKT transition to phase QLRO.  One can generalize to one-dimension, and consider a stack of
one--dimensional tubes of dipolar fermions \cite{1D_Dipolar_Bosons}. In this case, no phase
ordering can occur at any finite temperature, since the phase
dynamics are strictly one--dimensional. However, a dimerization
transition is still possible, leading to wider range of
dimerized, non-superfluid phase\cite{comment-1D}.
\par
    We expect that the Ising-XY model description of the layered
dipolar fermions will be insufficient deep in the BEC regime where
interaction energies are dominant compared to the Fermi-energy.
For sufficiently strong interactions or sufficiently dense
systems, the system will form a Wigner crystal \cite{Wigner_Crystal}.   Another possibility is that the formation of longer chains of three or more dipoles may become important \cite{Chains}.  In a regime where chains of $n$ dipoles are
favorable, a many-layered system would undergo $n$-merization
rather than dimerization.  Correspondingly, an $n$-merized phase may
undergo an $n$-state clock-model-type phase transition which generalizes
the Ising-type dimerization transition considered above.
Furthermore, for even $n$, bosonic chains could condense into an exotic
superfluid of dipolar chains.  Such states offer an intriguing
chance to examine the relatively unexplored boundary between
few-body interactions and many-body phase transitions, and deserve
further study.\par
    \textit{Acknowledgements -} We would like to acknowledge: E. Altman, T. Giamarchi, M. Lukin, D.F. Mross, D. Podolsky and S. Sachdev for helpful conversations.  This work was supported by: NSF IGERT Grant No. DGE-0801525, NSF grants DMR-0705472 and DMR-0757145,  AFOSR Quantum Simulation MURI, AFOSR MURI on Ultracold Molecules, DARPA OLE program, Harvard-MIT CUA, NSF grant DMR-09-06475.

\appendix
\section{Appendix A. BCS Gap Equation for Bilayer}
In this section, we show that the attractive component of intralayer dipolar interaction induces BCS pairing in a bilayer system.  We ignore repulsive intralayer interactions, as these serve only to renormalize the Fermi liquid parameters of the dipolar system.  The interaction between dipoles in adjacent layers separated by interlayer spacing $d$ along the $z$-axis, and by distance $r$ in the xy-plane, and with dipole moments $D$ polarized along the $z$-axis by an external field is:
\begin{equation} V_{\text{dip}}(r) = \frac{D^2}{\(r^2+d^2\)^{3/2}}\(1-\frac{3d^2}{r^2+d^2}\) \label{eq:DipPotRealSpace} \end{equation}
where we work in units with $4\pi\e_0=1$.  Fourier transforming (\ref{eq:DipPotRealSpace}) with respect to in-plane coordinate $r$, one finds:
\begin{equation} V^{(z,z+1)}_q = -D^2 qe^{-qd} \label{eq:DipPot}\end{equation}

We assume that pairing in the s-wave channel dominates, and that the transition temperature is set by condensation of Cooper pairs with zero center of mass momentum.  The self-consistency equation for the BCS order parameter $\Delta$ at temperature $T$ reads: \begin{eqnarray}\Delta_{z,\mathbf{k}} &=&-\sum_{\mathbf{k}'}V^{(z,z+1)}_{|\mathbf{k}-\mathbf{k}'|}\<\psi_{z+1,-\mathbf{k}'}\psi_{z,\mathbf{k}'}\> \nonumber \\
&=& -\frac{1}{2}\sum_{\mathbf{k}'}\frac{V^{(z,z+1)}_{|\mathbf{k}-\mathbf{k}'|} \Delta_{z,\mathbf{k}'}}{E_{k'}}\tanh\(\frac{E_k}{2T}\) \label{eq:BCSGapEqn}\end{eqnarray} 
In (\ref{eq:BCSGapEqn}), $E_k = \sqrt{\xi_{k}^2+\Delta_{z,\mathbf{k}}^2}$ where $\xi_k = \frac{k^2}{2m}-\mu$, $m$ is the effective mass, and $\mu$ is the chemical potential.  Since the number of particles on each layer is fixed, one must simultaneously solve for $\Delta$ using (\ref{eq:BCSGapEqn}) and for $\mu$ by fixing the particle density $n$:
\begin{equation} n = \sum_k \left[1-\frac{\xi_k}{E_k}\tanh\(\frac{E_k}{2T}\)\right] \label{eq:NumberConstraint}\end{equation}

\begin{figure}[hhh]
\begin{center}
\hspace{-.2in}
\includegraphics[width=3.5in]{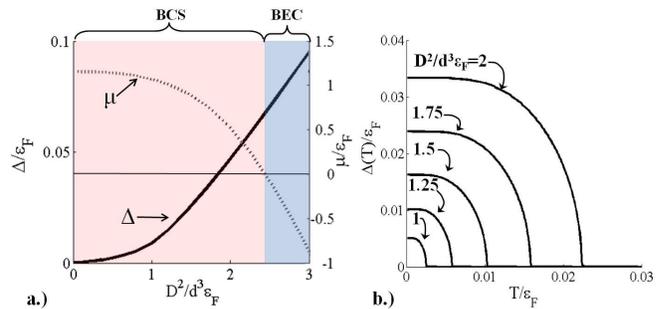}
\end{center}
\vspace{-.2in}
\caption{(color online) Numerical solutions of the BCS gap equation (\ref{eq:BCSGapEqn}) for the nearest-neighboring layer dipole potential (\ref{eq:DipPot}), at fixed particle density, and with $d/\lambda_F=1$, where $\lambda_F=2\pi/k_F$ is the Fermi wavelength.  (a) $T=0$ BCS pairing as a function of interaction strength $D^2/d^3$ (where energies are measured with respect to the Fermi energy $\e_F$ of the un-paired system).  Solid line shows solution for $\Delta$, and dotted line shows chemical potential $\mu$.  The BCS regime ($\mu>0$) and BEC regime ($\mu<0$) are highlighted in red and blue respectively.   (b) Temperature profile of the BCS gap for interaction strengths: $D^2/d^3\e_F =$ 2, 1.75, 1.5, 1.25, and 1 (highest to lowest respectively).  }
\label{fig:BCSGapProfile}
\end{figure}

We solve (\ref{eq:BCSGapEqn},\ref{eq:NumberConstraint}) numerically, using the expression (\ref{eq:DipPot}) for the dipole potential, and find a non-zero solution for any dipole moment $D$. Fig. (\ref{fig:BCSGapProfile})a. shows the $T=0$ solution for $\Delta$ as a function of interaction strength $D^2/d^3\e_F$, and Fig. (\ref{fig:BCSGapProfile})b. shows the temperature profile of the BCS gap for various interactions strengths.  These results demonstrate that intralayer interactions induce pairing for any value of dipole interaction strength.

In addition, one of us (D.W. Wang) has performed detailed studies of BCS pairing in $\mathcal{N} =$ 2, 3, and 4 layer systems.  These studies confirm that interlayer pairing occurs in these few-layered structures, and demonstrate explicitly that dimerization is favored for $\mathcal{N}=4$, but not for $\mathcal{N}=3$, in agreement with our analytic results from the Ginzburg-Landau free energy.  This work will be published elsewhere.

\section{Appendix B. Derivation of Ginzburg-Landau Free-Energy}
In this section, we include for completeness, details of the derivation of the Ginzburg-Landau (GL) free-energy (\ref{eq:GL}) from the microscopic fermionic action (\ref{eq:SFermionic}) for an $\mathcal{N}$ layer system.  This derivation, based on the Hubbard-Stratanovich (H-S) transformation, follows a standard route to deriving the GL free-energy in BCS theory (see for example A. Altland and B. Simons, Condensed Matter Field Theory (Cambridge University Press, Cambridge, 2006) p. 271-309).

Starting with (\ref{eq:GL}) we introduce H-S fields $\Delta_z(\mathbf{k}_\perp,\mathbf{Q}_\perp) = -\sum_{\mathbf{k}'}V^{(z,z+1)}_{|\mathbf{k}-\mathbf{k}'|}\<\psi_{z+1,-\mathbf{k}'+\mathbf{Q}/2}\psi_{z,\mathbf{k}'+\mathbf{Q}/2}\> $.  The subscript $z$ labels layer, and the two momentum labels $(\mathbf{k}_\perp,\mathbf{Q}_\perp)$ refer to the relative displacement and center of mass motion of the Cooper pairs respectively. The inclusion of $\mathbf{Q}_\perp\neq 0$ allows for spatial varations of the order parameters $\Delta_z$.  

With the introduction of the H-S fields $\Delta_z$, the fermionic action now reads: 
\begin{equation} S =  \sum_{z,k,k',Q} \Delta_{z,k,Q} \left(V^{-1}\right)_{k,k'} \Delta_{z,k',Q} + \sum_{k,Q} \Psi_{k,Q}^\dagger\mathcal{H}_\Delta\Psi_{k,Q}
\end{equation}
\begin{equation}
\mathcal{H}_\Delta =
\begin{pmatrix} i\omega+\xi& \Delta_{1,k,Q} & 0  & \cdots \\
                \bar\Delta_{1,k,Q} & i\omega-\xi& -\bar\Delta_{2,-k,Q}&\cdots\\
					0&-\Delta_{2,k,Q}&i\omega+\xi&\cdots\\
					\vdots&\vdots&\vdots&\ddots\end{pmatrix}
\label{eq:H_Delta}
\end{equation}
where $\Psi_{k,Q}\equiv\begin{pmatrix} \bar\psi_{1,-k+Q} & \psi_{2,k} & \bar\psi_{3,-k+Q}&\psi_{4,k} & \hdots \end{pmatrix}^T$, and $\omega$ is an (imaginary) Matsubara frequency label.  Here, $(V^{-1})_{k,k'}$ denotes $(k,k')$ component of the inverse (in the operator sense) of the dipole potential (\ref{eq:DipPot}).  As will be explained below, it is not necessary to explicitly compute this inverse in order to develop the GL theory.  Since we are concerned with finite temperature phase transitions, we neglect quantum fluctuations by treating $\Delta$ as independent of $\omega$.  This corresponds to developing the GL free-energy for the $\omega=0$ component of $\Delta_z$.  We also assume that the relative displacement momentum $k$ profile of $\Delta_{k,Q}$ does not fluctuate substantially from the mean-field form.  This corresponds to fixing the form of the Cooper pair wave-function to the one most energetically favored at the mean-field level, and is justified by the fact that the dominant instability towards pairing will occur with this pairing profile.  With this assumption, the $k$ labels on $\Delta$ are non-dynamical and will be dropped from subsequent expressions.  We have validated this approach by checking that the resulting GL free-energy derived in this way can accurately reproduce the results of the numerical solution to the BCS gap-equation. 

Integrating out the Fermions gives the following effective action for the Hubbard-Stratonovich fields $\Delta_z$:
\begin{widetext}
\begin{eqnarray} S[\Delta] &=& \sum_n S^{(2n)}[\Delta] = \text{Tr}\ln\left[1+G_0^{-1}\mathcal{H}_\Delta\right]=
\nonumber \\
&=& \sum_{n=1}^\infty \frac{1}{2n}Tr\left[\begin{pmatrix} G_0^h&0&0&\cdots\\0&G_0^p&0&\cdots\\0&0&G_0^h&\cdots\\ \vdots&\vdots&\vdots&\ddots\end{pmatrix} \begin{pmatrix}0&\Delta_{1}(Q)&0&\cdots\\\bar\Delta_{1}(Q)&0&-\bar\Delta_{2}(Q)&\cdots\\0&-\Delta_{2}(Q)&0&\cdots\\
 \vdots&\vdots&\vdots&\ddots\end{pmatrix} \right]^{2n}
\nonumber\\
&\hspace{-.3in}=&\sum_{n=1}^\infty \frac{1}{2n}Tr\left[\begin{pmatrix}0&\Delta_1&0&0&\cdots&0&0\\ \bar\Delta_1&0&-\bar\Delta_2&0&\cdots&0&0\\ 0&-\Delta_2&0&\Delta_3&\cdots&0&0\\ 0&0&\bar\Delta_3&0&\cdots&0&0\\ \vdots&\vdots&\vdots&\vdots&\ddots&\vdots&\vdots\\ 0&0&0&0&\cdots&0&\Delta_\mathcal{N}\\ 0&0&0&0&\cdots&\bar\Delta_\mathcal{N}&0  \end{pmatrix}^2 
G_{0,k}^pG_{0,-k+Q}^h\right]^{n}
\nonumber\\
&\hspace{-.3in}=&\sum_{n=1}^\infty \frac{1}{2n}Tr\left[\begin{pmatrix}|\Delta_1|^2&0&-\Delta_1\bar\Delta_2&0&\cdots&0&0\\  0&|\Delta_1|^2+|\Delta_2|^2&0&-\bar\Delta_2\Delta_3&\cdots&0&0\\ -\Delta_2\bar\Delta_1&0&|\Delta_2|^2+|\Delta_3|^2&0&\cdots&0&0\\  \vdots&\vdots&\vdots&\vdots&\ddots&\vdots&\vdots\\ 0&0&0&0&\cdots&|\Delta_{\mathcal{N}-2}|^2+|\Delta_{\mathcal{N}-1}|^2&0\\ 0&0&0&0&\cdots&0&|\Delta_N|^2  \end{pmatrix} G_{0k}^pG_{0,-k+Q}^h\right]^{n}\nonumber \\
\label{eq:FermionDet}\end{eqnarray}
\end{widetext}
where we have dropped all irrelevant constant terms that do not depend on $\Delta$, and $G_{0,k}^p = \frac{1}{i\omega-\xi_k}$, $G_{0,k}^h = \frac{1}{i\omega+\xi_k}$ are the particle and hole Green functions respectively (note the precise form of the bottom right entry of the second line of (\ref{eq:FermionDet}) depends on whether $\mathcal{N}$ is odd or even).  Also, in (\ref{eq:FermionDet}) momentum labels on $\Delta$ and $G_0$ have been suppressed where possible in order to conserve space.  

We truncate the above series at quartic order, which is formally justified near the pairing transition temperature $T_c^{\text{MF}}$ where $\Delta$ is small.  The GL coefficients $\{\kappa,r,u\}$ can then be computed by explicitly evaluating the trace over the products of fermion Green functions appearing in (\ref{eq:FermionDet}).  The terms quadratic in $\Delta$ are: $S^{(2)} = \sum_q \Gamma_Q^{-1}|\Delta(Q)|^2$ where $\Gamma(Q,T) \equiv V^{-1}-\frac{T}{\Omega}\sum_p G^p_{0,p}G^h_{0,-p+Q}$, and where $\Omega$ is the system volume.  The mass term, with coefficient $r$, determines the energy of having a uniform pairing amplitude $|\Delta|$.  To compute $r$, we note that at $T_c^{\text{MF}}$ we have: $\Gamma(0,T_c^{\text{MF}}) = 0=V^{-1}-\frac{T_c^{\text{MF}}}{\Omega}\sum_p G^{p}_{0,p}G^{h}_{0,-p}$, indicating that for $T$ near $T_c^{\text{MF}}$ one can expand:
\begin{eqnarray}r  &=&(T-T_c^{\text{MF}})\frac{-\partial}{\partial T} \sum_p G^{p}_{0,p}G^{h}_{0,-p} \nonumber \\
&=& (T-T_c^{\text{MF}})\int \frac{d^2p}{(2\pi)^2}\frac{-\partial_T n_F(\xi_p,T_c^{\text{MF}}) }{\xi_p}
\nonumber \\
 &=&  \nu t\end{eqnarray}
where $n_F(\e,T)$ is the Fermi-distribution at energy $\e-\mu$ and temperature $T$.    The coefficient $\kappa$, of the gradient term $\kappa |\nabla_\perp \Delta|^2$, is obtained by expanding $\Gamma(Q,T)$ to quadratic order in $Q$: 
\begin{eqnarray}\kappa &=& \frac{\partial}{\partial Q^2}\left[-\frac{T}{\Omega}\sum_p G_{0,p}G_{0,-p+Q}\right] \\
&=& \frac{\partial}{\partial Q^2}\int \frac{d^2p}{(2\pi)^2}\(\frac{\mathbf{Q}\cdot\mathbf{p}}{2m}\)^2\frac{-\partial_\xi^2n_F(\xi_p,T)}{2\xi_p}= \frac{7\zeta(3)}{32\pi^3}\frac{\e_F}{T^2}\nonumber \end{eqnarray}
Note that obtaining $\kappa$ and $r$ by expanding $\Gamma(Q,T)$ to leading order in $Q$ and $T$ respectively has allowed us to neatly sidestep the explicit computation of $V^{-1}$. Anothering interesting feature is that the details of the dipole potential are fully contained in the single parameter, $T_c^{\text{MF}}$, which we obtain by numerically solving (\ref{eq:BCSGapEqn}).  

Finally, we turn to the evaluation of the quartic term coefficient $u$, which comes from terms of the form $S^{(4)} \sim |\Delta|^4\sum_p \(G^{p}_pG^{h}_{-p}\)^2$.  The Matsubara frequency summation in $\sum_p \(G^{p}_pG^{h}_{-p}\)^2$ can be done explicitly:
\begin{eqnarray} &\sum_{\omega_n}&\frac{1}{(i\omega_n-\xi_p)^2}\frac{1}{(i\omega_n+\xi_p)^2}
\nonumber\\
 &=& \oint \frac{-dz}{2\pi i}\frac{\beta}{e^{\beta z}+1}\frac{1}{(z-\xi_p)^2}\frac{1}{(z+\xi_p)^2}\nonumber\\
&=& \frac{-\beta^2}{8\xi_p^2\cosh^2(\beta\xi_p/2)}+\frac{\beta}{4\xi_p^3}\tanh(\beta\xi_p/2)
\end{eqnarray}
where $\beta = 1/T$ is the inverse temperature.  The remaining integral over $\vec p$ can be rendered dimensionless and computed numerically yielding:
\begin{equation} \kappa = \frac{\nu \beta^3}{32}\int_{-\infty}^\infty dx\left(-\frac{1}{x^2\cosh^2x}+\frac{\tanh x}{x^3}\right) = \frac{1.7}{32}\frac{\nu}{T^3}\end{equation} 
Using the above computations of $\{\kappa,r,u\}$ it is now a straightforward matter to evaluate (\ref{eq:FermionDet}), which for infinite number of layers gives:
\begin{eqnarray} F &=& \sum_z \int d^2r \Big{(} \kappa|\nabla_{\hspace{-.05in}\perp}\Delta_z|^2+r|\Delta_z|^2 \nonumber \\ &&\hspace{.5in}+u|\Delta_z|^4+2u|\Delta_z|^2|\Delta_{z+1}|^2\Big{)} \label{eq:AppGL}\end{eqnarray}
The relative factor of $2$ between the $|\Delta_z|^4$ and $|\Delta_z|^2|\Delta_{z+1}|^2$ terms can be obtained by a careful accounting of combinatorics.  Alternatively, this factor may be checked by considering a trilayer system whose microscopic $SU(2)$ symmetry (described in the main text) guarantees that the quartic terms in the GL free energy be of the form: $S^{(4)}_{\mathcal{N}=3}=u(|\Delta_1|^2+\Delta_2|^2)^2$.

\section{Appendix C. Coupling Between Domain Walls and Phase Fluctuations}
In this section, we discuss some subtleties involved in coarse graining the GL free-energy (\ref{eq:GL}) to arrive at the effective lattice model (\ref{eq:LatticeModel}).  In deriving the value of the XY stiffness $J(\sigma_i,\sigma_j)$, which couples the dimerization order parameter to phase fluctuations, we have made the simplifying approximation that the the superfluid stifness drops abruptly to zero at the location of an in-plane Ising domain wall.  

In reality, there is some residual superfluid stiffness that varies continuously across the length of the domain wall.  Keeping track of this residual stiffness corresponds to including higher order gradient terms such as $\kappa^{(2)}|\nabla_\perp \Delta_z(r)|^4$, which includes terms of the form $|\Delta|^2|\nabla_\perp\Delta(r)|^2|\nabla\theta(r)|^2$ that couple domain walls and phase fluctuations.  However, these terms are subleading and have a negligible effect on the systems phase diagrams.  Formally, this is because such higher order gradient terms are highly irrelevant in the renormalization group sense.  

We have also conducted separate simulations that include residual phase stiffness at the dimerization domain boundaries, and have explicitly confirmed that the resulting phase diagram is highly insensitive to the inclusion of such terms.  Specifically, in addition to the usual $J(\sigma_i = +1,\sigma_j = +1) = J_0$ and $J(-1,-1)=0$, we allowed for $J(+1,-1) = J(-1,+1) = J_{\text{res}}$, and verified that the choice of value for $J_{\text{res}}$ had little discernible effect on either the dimerization or BKT phase transitions.

A second simplifying assumption is used in parameterizing the in-plane domain wall as $\Delta(x) = \frac{|\Delta_0|}{2}(1+\alpha(x))$. This parameterization implicitly assumes constant phase over the range of the domain wall, which is justified because $\ell_{\mathrm{DW}}\sim \xi_{\mathrm{GL}}$. Since $\xi_{\mathrm{GL}}$ is the shortest scale over which the phase is well-defined, it is essentially constant on lengthscales $\leq \xi_{\mathrm{GL}}$.



\begin{thebibliography}{99}
\bibitem{comment-1D}
In 1D, where the Fermi-surface is perfectly nested, one must also consider possible charge ordered states.
\end{thebibliography}
\end{document}